\documentclass[prd,aps,groupedaddress,
showpacs,
twocolumn,%
nofootinbib,superscriptaddress
]{revtex4}%
\usepackage{graphicx}
\usepackage{amsmath}
\usepackage{bm}
\usepackage{epsfig}
\usepackage{amssymb}
\begin{document}
\title{Polarization of prompt 
$\mbox{\boldmath$J/\psi$}$ in
$\mbox{\boldmath$pp \to J/\psi + X$}$ 
at 
$\mbox{\boldmath$\sqrt{s}=200\,\textrm{GeV}$}$ 
}
\author{Hee Sok Chung}
\affiliation{Department of Physics, Korea University, Seoul 136-701, Korea}
\author{Seyong Kim}
\affiliation{Department of Physics, Sejong University, Seoul 143-747, Korea}
\affiliation{School of Physics, Korea Institute for Advanced Study,
  Seoul 130-722, Korea}
\author{Jungil Lee}
\affiliation{Department of Physics, Korea University, Seoul 136-701, Korea}
\author{Chaehyun Yu}
\affiliation{School of Physics, Korea Institute for Advanced Study,
  Seoul 130-722, Korea}

\date{\today}
\preprint{}
\begin{abstract}
Within the framework of the nonrelativistic QCD factorization
approach, we compute the cross section and polarization of prompt $J/\psi$ 
produced from proton-proton collisions at the center-of-momentum energy
$\sqrt{s}=200\,$GeV. 
We present the transverse-momentum distribution 
in the forward-rapidity region $1.2<|y|<2.2$
and the rapidity distribution
over the transverse-momentum range $2\,\textrm{GeV}<p_T<20\,\textrm{GeV}$.
The perturbative contributions are computed
at leading order in the strong coupling constant.
We predict slight transverse polarization of $J/\psi$ in the forward-rapidity
region, while that for the midrapidity region is slightly longitudinal.
The transverse-momentum distribution agrees well with the PHENIX preliminary
data and the color-singlet-model prediction at next-to-leading order in
$\alpha_s$, but disagrees with the result from the leading-order color-singlet
model or the s-channel-cut method.
\end{abstract}

\maketitle
\section{Introduction}
As one of the long-standing puzzles in heavy-quarkonium phenomenology,
the polarization of prompt $J/\psi$ produced with large transverse
momentum ($p_T$) at hadron colliders has not been clearly understood yet. 
According to the nonrelativistic quantum chromodynamics (NRQCD)
factorization approach \cite{Bodwin:1994jh}, the dominant production
mechanism of $J/\psi$ with large $p_T$ in hadron collisions
is the gluon fragmentation into the color-octet spin-triplet $S$-wave
heavy-quark-antiquark pair [$Q\bar{Q}_8({}^3S_1)$] that evolves into the
meson~\cite{Braaten:1994vv}. Because the corresponding long-distance
transition preserves the spin at leading order in $v$, which is
the typical velocity of the heavy quark of the meson, the $J/\psi$ should
be transverse at large $p_T$ \cite{Cho:1994ih}. However, the run II
measurement by the Collider Detector at Fermilab (CDF)
Collaboration \cite{Abulencia:2007us} agrees with neither the run I 
data \cite{Affolder:2000nn} nor the NRQCD
predictions~\cite{Beneke:1995yb,Beneke:1996yw,%
Leibovich:1996pa,Braaten:1999qk,Kniehl:2000nn}, although the feeddowns
from the $P$-wave spin-triplet states $\chi_{cJ}$
for $J=0$, $1$, and $2$ dilute the transverse
polarization~\cite{Braaten:1999qk,Kniehl:2000nn}. The problem has not
been resolved even after considering corrections in various 
aspects~
\cite{Campbell:2007ws,Artoisenet:2007xi,Gong:2008sn,Gong:2008hk,%
Artoisenet:2008zz,Gong:2008ft,Fan:2009zq,Gong:2010bk,Ma:2010yw,%
Butenschoen:2010rq,Butenschoen:2010px,Ma:2010jj}. 
Therefore, it is worthwhile
to see if NRQCD still confronts with experiments of lower center-of-momentum
(CM) energies $\sqrt{s}$, where the dominance of the gluon fragmentation
may not happen.

In this aspect, the \textit{inclusive} $J/\psi$ production in $p p$
collisions at $\sqrt{s}=200\,$GeV,
where the fragmentation does not dominate~\cite{Cooper:2004qe},
at the Brookhaven Relativistic Heavy-Ion Collider (RHIC) may
provide independent experimental constraints to the $J/\psi$ polarization.
Because only a tiny fraction of the PHENIX inclusive $J/\psi$
samples contain the $J/\psi$ from the $B$ decay \cite{Oda:2008zz}, we can
treat this as prompt $J/\psi$. In the case of the cross section, the NRQCD
predictions in Refs.~\cite{Nayak:2003jp, Cooper:2004qe} are consistent with
the run-3 PHENIX data~\cite{Adler:2003qs,deCassagnac:2004kb}. As a new
production mechanism, the $s$-channel-cut approach~\cite{Haberzettl:2007kj},
which suffers from a criticism~\cite{Artoisenet:2009mk}, also explains
the data~\cite{Lansberg:2005pc}. The color-singlet-model (CSM) prediction for
the direct $J/\psi$ production rate has been computed at next-to-leading
order (NLO) in the strong coupling $\alpha_s$~\cite{Brodsky:2009cf,%
Lansberg:2010vq}, 
including contributions from $cg$ fusion. The NLO CSM prediction 
agrees with the data for the rapidity distribution, 
assuming that about $40\,\%$ of
the $J/\psi$ samples are from higher resonances,
while it underestimates the data for the $p_T$ distribution.
The NLO NRQCD prediction has been found to agree with the 
data for the $p_T$ distribution~\cite{Butenschoen:2010rq, Butenschoen:2010px}.
In addition to the cross section~\cite{Adare:2006kf}, the PHENIX Collaboration 
has measured the polarization~\cite{Adare:2009js} of the inclusive $J/\psi$.
In the midrapidity region $|y|<0.35$, the PHENIX data for the $J/\psi$
polarization~\cite{Adare:2009js} agree with the predictions
of both NRQCD \cite{Chung:2009xr} and 
the s-channel-cut method~\cite{Lansberg:2008jn}.
However, the PHENIX preliminary data in Ref.~\cite{daSilva:2009yy} for
the forward-rapidity region $1.2<|y|<2.2$ disfavor the s-channel-cut
prediction~\cite{Lansberg:2008jn,daSilva:2009yy}.
The PHENIX Collaboration is currently carrying out a comprehensive
analysis with the data from the forward-rapidity region.
Therefore, it is desirable to provide NRQCD predictions that can be compared
with the forthcoming updated PHENIX data for the inclusive $J/\psi$
polarization in the forward-rapidity region.

In this paper, we present an updated NRQCD prediction for the
polarization of
prompt $J/\psi$ produced in the forward-rapidity region $1.2<|y|<2.2$
in $pp$ collisions at $\sqrt{s}=200\,$GeV. The perturbative contributions
are computed at LO in $\alpha_s$, like the previous analysis in
Ref.~\cite{Chung:2009xr}, in which the cross section and polarization in 
the midrapidity region $|y|<0.35$ within the $p_T$ range
$1.5\,\textrm{GeV}<p_T<5\,\textrm{GeV}$ are predicted in detail. 
The PHENIX Collaboration is also preparing to release the rapidity dependence
of the $J/\psi$ in both the midrapidity and forward-rapidity regions. In this
ongoing analysis, they are using a conservative lower $p_T$ cut
$p_T>2\,\textrm{GeV}$ to avoid the region where LO NRQCD may break down.
We also present a NRQCD prediction for the rapidity dependence of the 
polarization of prompt $J/\psi$ in the region $|y| < 3$.
Except for these changes in the kinematic region, 
our analysis presented here is very similar to that in Ref.~\cite{Chung:2009xr}
so we do not reproduce details of our strategy to compute observables
and explanations on the input parameters given in Ref.~\cite{Chung:2009xr}. 

The remainder of this paper is organized as follows. We first present
the NRQCD prediction for the $p_T$ dependence of the differential cross
section followed by the NRQCD prediction for the $p_T$ dependence of
the polarization parameter $\alpha$ integrated over the forward-rapidity
region in Sec.~\ref{sec:pt}. The rapidity distributions of the cross
section and polarization parameter $\alpha$ integrated over the $p_T$ range 
$2\,\textrm{GeV}<p_T<20\,\textrm{GeV}$ are given in Sec.~\ref{sec:y}, and
we summarize in Sec.~\ref{sec:summary}.
\section{$\bm{p_T}$ distributions\label{sec:pt}}
\begin{figure}
\epsfig{file=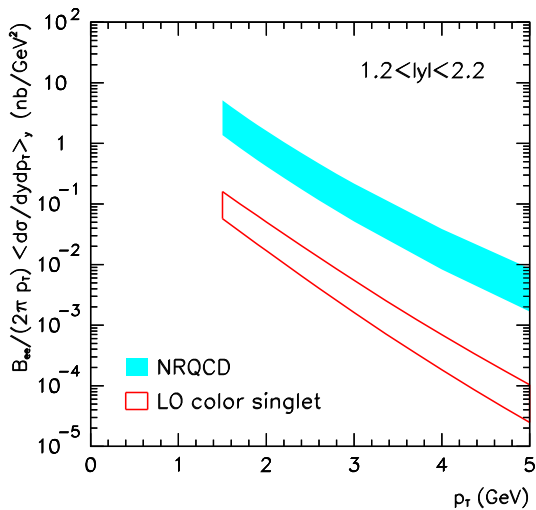,width=7cm}
\caption{The differential cross section of prompt $J/\psi$
with the rapidity $1.2 < | y | < 2.2$ 
in $pp$ collisions at $\sqrt{s}=200\,$GeV
as a function of $p_T$.
The shaded band represents the NRQCD prediction and the band
surrounded by a solid curve is the color-singlet contribution
at LO in $\alpha_s$.  
\label{fig:sigma1}
}
\end{figure}
\begin{figure}
\epsfig{file=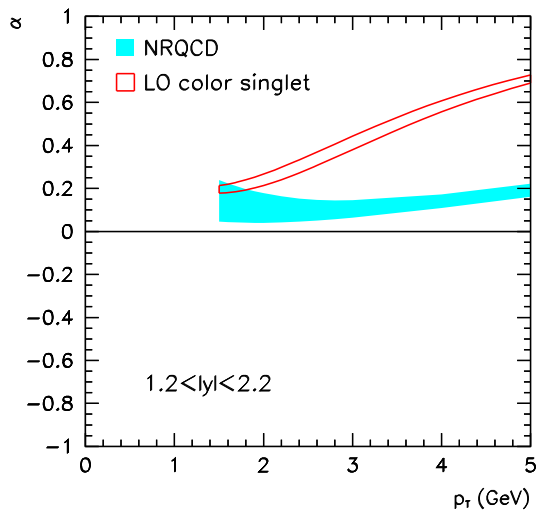,width=7cm}
\caption{The polarization parameter $\alpha$ for the
prompt $J/\psi$ 
with the rapidity $1.2 < | y | < 2.2$ 
in $pp$ collisions at $\sqrt{s}=200\,$GeV
as a function of $p_T$.  \label{fig:alpha1}}
\end{figure}
In this section, we provide the NRQCD predictions for the $p_T$
distributions of the production rate and polarization parameter $\alpha$
for the prompt $J/\psi$ produced in $pp$ collisions at $\sqrt{s}=200$\,GeV.
The basic strategy to compute the cross section and polarization parameter
$\alpha$ can be found in Ref.~\cite{Chung:2009xr}.

The cross section for the prompt $J/\psi$ in $pp$ collisions
is computed by including the LO parton
processes $ij\to c\bar{c}+k$, with $i,\,j=g,\,q,\,\bar{q}$ and $q=u,\,d,\,s$.
We use numerical values for the NRQCD matrix elements (ME) 
given in Ref.~\cite{Braaten:1999qk}. For the parton
distribution functions (PDF), we choose MRST98LO~\cite{Martin:1998sq} as
the default value and CTEQ5L~\cite{Lai:1999wy} for comparison. 
The transverse mass $m_T=(4 m_c^2 + p_T^2)^{1/2}$ 
is used for both the factorization and renormalization scales $\mu$
with the charm-quark mass $m_c=1.5$ GeV. $\alpha_s$ is evaluated
from the one-loop formula using the value of $\Lambda_{\rm QCD}$ given
in each PDF set~\cite{Martin:1998sq,Lai:1999wy}. 
To estimate theoretical uncertainties, we follow the method given in
Ref.~\cite{Chung:2009xr}.

Our prediction for the differential cross section of
prompt $J/\psi$ is shown in 
Fig.~\ref{fig:sigma1} as a function of $p_T$.
The shaded band indicates the LO NRQCD prediction and the LO CSM 
contribution is displayed as a band surrounded by a solid curve. 
The rate is averaged over the forward-rapidity range $1.2<|y|<2.2$,
and its explicit normalization is given by
\begin{equation*}
\label{dsdy-norm}
\frac{B_{ee}}{2 \pi p_T} \langle d^2 \sigma/dy dp_T \rangle_y
= \frac{B_{ee}}{2 \pi p_T \times 2} 
\int_{1.2<|y|<2.2} \frac{d ^2 \sigma}{dy dp_T} dy,
\end{equation*}
where $B_{ee}$ is the branching fraction for $J/\psi\to e^+e^-$ and
the factor $2$ in the denominator on the right side
is from the range of integration  $1.2<|y|<2.2$.

The result shows that more than $70\,\%$ of the prompt $J/\psi$'s are direct: 
At $p_T = 1.5\,$GeV ($5\,$GeV), the contributions from
direct $J/\psi$, feeddown from $\psi(2S)$, and feeddown from $\chi_{cJ}$
are $83\,\%$ ($71\,\%)$, $8\,\%$ ($11\,\%$), and $10\,\%$ ($19\,\%$) of 
the prompt $J/\psi$'s, respectively.
According to Fig.~\ref{fig:sigma1}, the LO color-singlet 
${}^3S_1$ contribution is negligible, which amounts to 
$4\,\%$ ($2\,\%$) of the direct $J/\psi$, while the color-octet 
${}^3P_J$ or ${}^1S_0$ and ${}^3S_1$ channels are
$94\,\%$ ($81\,\%$) and $1.5\,\%$ ($17\,\%$), respectively,
at $p_T = 1.5\,$GeV ($5\,$GeV). 
The uncertainties are evaluated by considering the variations of 
$m_c=1.50\pm 0.05$\,GeV,
$m_T/2\le \mu\le 2m_T$, NRQCD ME,
and PDF \cite{Martin:1998sq,Lai:1999wy}
as described in Ref.~\cite{Chung:2009xr}.
The uncertainties from $\mu$ of $83\,\%$ ($98\,\%$)
dominate over those from $m_c$
[$10\,\%$ ($2\,\%$)] and those from the NRQCD ME and 
PDF [$7\%$ ($0\%$)] at $p_T = 1.5\,$GeV ($5\,$GeV).

The polarization parameter 
$\alpha$ is defined by 
$\alpha = (\sigma_T-2\sigma_L)/(\sigma_T+2\sigma_L)$,
where $\sigma_T$ ($\sigma_L$) is the cross section for the 
transversely (longitudinally) polarized $J/\psi$.
If $J/\psi$ is completely transverse (longitudinal),
then $\alpha=+1\,(-1)$. If $J/\psi$ unpolarized, then $\alpha=0$.
The polarized cross sections depend on the frame.
We choose the hadron CM frame that was employed in the PHENIX analysis.

The $p_T$ dependence of the polarization parameter $\alpha$ integrated over
the forward-rapidity region $1.2<|y| < 2.2$ is shown in Fig.~\ref{fig:alpha1}
with the same style as Fig.~\ref{fig:sigma1}.
The NRQCD prediction is slightly transverse over the whole range of $p_T$,
while the color-singlet contribution gets more transverse as $p_T$
increases. In comparison with the prediction for the midrapidity region,
the uncertainties mostly ($\ge 98\%$) come from the NRQCD ME and PDF.

Our results for the polarization agree with the PHENIX preliminary
data \cite{daSilva:2009yy} and are also compatible with the CSM
prediction at NLO \cite{Lansberg:2010vq}, which has larger uncertainties
than the NRQCD prediction. The s-channel-cut prediction disagrees with
both NRQCD and the PHENIX preliminary data \cite{daSilva:2009yy}. 
\section{Rapidity distributions\label{sec:y}}
\begin{figure}
\epsfig{file=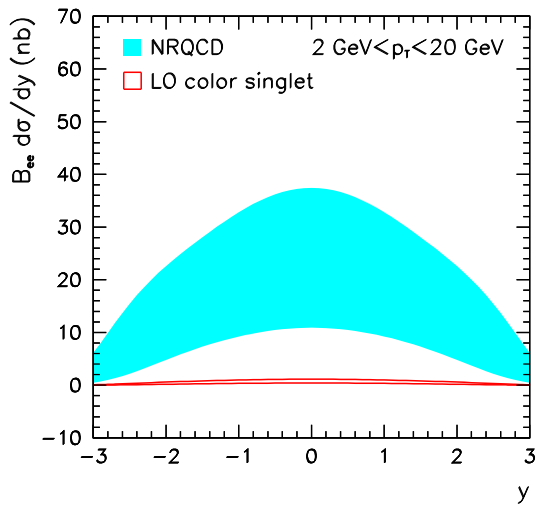,width=7cm}
\caption{The differential cross section of prompt $J/\psi$
with the transverse momentum $2\,\textrm{GeV}<p_T<20\,\textrm{GeV}$
in $pp$ collisions at $\sqrt{s}=200\,$GeV
as a function of $y$.
\label{fig:sigma2}
}
\end{figure}

\begin{figure}
\epsfig{file=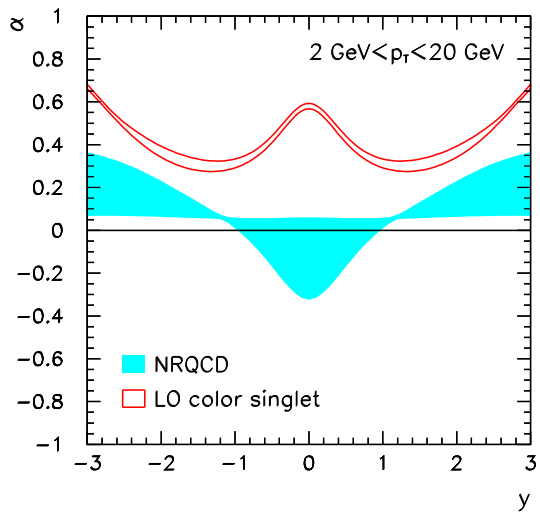,width=7cm}
\caption{The polarization parameter $\alpha$ for the prompt $J/\psi$ 
with the transverse momentum $2\,\textrm{GeV}<p_T<20\,\textrm{GeV}$
in $pp$ collisions at $\sqrt{s}=200\,$GeV
as a function of $y$.
\label{fig:alpha2}}
\end{figure}
In this section, we provide the NRQCD predictions for the rapidity
distributions of the production rate and polarization parameter $\alpha$.
The PHENIX analysis of the rapidity dependence of the $J/\psi$ is
under way with a conservative cut $p_T>2\,\textrm{GeV}$ which avoids
the kinematic region where the fixed-order calculations may break down.
Because the differential cross section at $p_T = 5\,(20)\,\textrm{GeV}$
is less than $10 \,\%$ $(0.1\,\%)$ of that for $p_T = 2\,\textrm{GeV}$,
the total cross section should be insensitive to an
upper limit $\agt 10\,\textrm{GeV}$ and the neglect of the gluon
fragmentation can be justified. Our results show that about 79\,\% of
the total production rate comes from the direct $J/\psi$, and feeddowns 
from $\chi_{cJ}$ and $\psi(2S)$ occupy 13\,\% and 8\,\%, respectively.
In Fig.~\ref{fig:sigma2}, we show the NRQCD prediction for the
differential cross section integrated over the  range 
$2\,\textrm{GeV}<p_T<20\,\textrm{GeV}$ as a function of $y$.
According to this figure, the LO color-singlet contribution is negligible
over the rapidity range $|y|<3$.
The lower cut for $p_T$ severely eliminates the color-octet and color-singlet
${}^3S_1$ contributions to occupy only a few percent of the total production
rate, making ${}^3P_J$ and ${}^1S_0$ octet channels dominate.
The dominant source of the uncertainties in the rapidity distribution
is the scale dependence which occupies about $92\,\%$ ($96\,\%$) 
at $y=0$ ($|y|= 3$). The contribution of $m_c$ dependence is about
$6\,\%$ ($3\,\%$) and that of NRQCD ME and PDF is about $3\,\%$
($1\,\%$) at $y=0$ ($|y|=3$).

The rapidity distribution of $\alpha$ integrated over the
range $2\,\textrm{GeV}<p_T<20\,\textrm{GeV}$ 
is shown in Fig.~\ref{fig:alpha2}.
The LO CSM prediction is always transverse 
over the whole range of the rapidity.
The central values of the NRQCD prediction are slightly longitudinal at
midrapidity region, and there is a turnover around $|y|\approx 1$ from which
the NRQCD prediction becomes transverse. The uncertainties of the NRQCD
prediction become maximum (minimum) at $y=0$ ($|y|\approx 1$). Therefore,
the uncertainties of the NRQCD prediction for the forward-rapidity region
in Fig.~\ref{fig:alpha1} are significantly smaller than those for the
midrapidity region presented in Fig.~2 of Ref.~\cite{Chung:2009xr}.
More than $96\%$ of the uncertainties of the predictions in 
Fig.~\ref{fig:alpha1} are from the NRQCD ME
and PDF. The uncertainties from $\mu$ and $m_c$ are negligible.
\section{Summary\label{sec:summary}}
In summary, we have presented the NRQCD predictions for 
the $p_T$ and $y$ distributions of the differential cross section 
and the polarization parameter $\alpha$
for the prompt $J/\psi$ produced in $p p$ collisions at
$\sqrt{s}=200\,$GeV. 
The $p_T$ and $y$ distributions are given in the kinematic ranges
$1.2<|y|<2.2$ and $2\,\textrm{GeV}<p_T<20\,\textrm{GeV}$,
respectively.
While the NRQCD prediction for the prompt $J/\psi$ polarization
in the midrapidity region ($|y|<0.35$) is slightly longitudinal,
that in the forward-rapidity region ($1.2<|y|<2.2$) is slightly
transverse with less theoretical uncertainties.
The change in the lower-$p_T$ cut to $2\,\textrm{GeV}$
also reduced the uncertainties in our prediction by avoiding
a possible break down of fixed-order calculations at low $p_T$.
The LO CSM prediction severely underestimates the differential cross
section and shows more transverse $J/\psi$ than that of NRQCD.
Our polarization prediction is less transverse as that of the 
s-channel-cut method~\cite{Lansberg:2008jn} ($\alpha=0.36$),
but is compatible with the NLO CSM prediction~\cite{Lansberg:2010vq},
which has larger uncertainties than our results. We anticipate our
NRQCD predictions presented here can be tested against a forthcoming
updated analysis by the PHENIX Collaboration.
\begin{acknowledgments}
We express our gratitude to Marzia Rosati and Cesar Luiz da Silva 
for drawing our attention to the problem discussed in this paper.
This work was
supported by Basic Science Research Program through
the NRF of Korea funded by the MEST under Contracts
No. 2010-0027811 (H.S.C.), No. 2010-0028228 (J.L.), and No. 2009-0072689 (C.Y.).
SK is supported by the International Research and Development Program of 
the NRF of Korea funded by the MEST (Grant No. K20803011439-10B1301-01410).
\end{acknowledgments}


\end{document}